\documentclass[preprintnumbers]{revtex4-2}
\usepackage{dcolumn}
\usepackage{bm}
\usepackage{graphicx}
\usepackage{amsmath}

\begin{document}

\title{Semi-vortex solitons and their excited states in spin-orbit-coupled
binary bosonic condensates}
\author{Haiming Deng$^{1,2,3}$, \ Jinqing Li$^{1,2,3}$, Zhaopin Chen$^{4}$,
Yaohui Liu$^{3}$, \ Dong Liu$^{3}$, \ Chunzhi Jiang$^{1,2,3}$, \ Chao Kong$%
^{1,2,3}$\thanks{%
Corresponding Author: superkong@xnu.edu.cn}, and Boris A. Malomed$^{5,6}$}
\address{$^1$ School of Physics and Electronic-Electrical Engineering, Xiangnan
University, Chenzhou 423000, China\\
$^2$ Microelectronics and Optoelectronics Technology Key Laboratory of Hunan
Higher Education, Xiangnan University, Chenzhou 423000, China\\
$^3$ Hunan Engineering Research Center of Advanced Embedded Computing and
Intelligent Medical Systems, Xiangnan University, Chenzhou 423000, China\\
$^{4}$Physics Department and Solid-State Institute, Technion, Haifa 32000,
Israel\\
$^{5}$Department of Physical Electronics, School of Electrical Engineering,
Faculty of Engineering, Tel Aviv University, Tel Aviv 69978, Israel\\
$^{6}$Instituto de Alta Investigaci\'{o}n, Universidad de Tarapac\'{a},
Casilla 7D, Arica, Chile}

\begin{abstract}
It is known that two-dimensional two-component fundamental solitons of the
\textit{semi-vortex} (SV) type, with vorticities $\left( s_{+},s_{-}\right)
=\left( 0,1\right) $ in their components, are stable ground states (GSs) in
the spin-orbit-coupled (SOC) binary Bose-Einstein condensate with the
contact self-attraction acting in both components, in spite of the
possibility of the critical collapse in the system. However, excited states
(ESs) of the SV solitons, with the vorticity set $\left( s_{+},s_{-}\right)
=\left( S_{+},S_{+}+1\right) $ and $S_{+}=1,2,3,...$, are unstable in the
same system. We construct ESs of SV solitons in the SOC system with opposite
signs of the self-interaction in the two components. The main finding is%
\emph{\ stability} of the ES-SV solitons, with the extra vorticity (at
least) up to $S_{+}=6$. The threshold value of the norm for the onset of the
critical collapse, $N_{\mathrm{thr}}$, in these excited states is\emph{\
higher} than the commonly known critical value, $N_{c}\approx 5.85$,
associated with the single-component Townes solitons, $N_{\mathrm{thr}}$
increasing with the growth of $S_{+}$. A velocity interval for stable motion
of the GS-SV solitons is found too. The results suggest a solution for the
challenging problem of the creation of stable vortex solitons with high
topological charges.

\end{abstract}

\maketitle

\section{Introduction}

The experimental creation of Bose-Einstein condensates (BECs) in ultracold
atomic gases \cite{MHAnderson,CCBradley,KBDavis} provides an ideal platform
for exploring various states in quantum superfluids, such as vortices \cite%
{JEWilliams,Kasamatsu}, stable dark \cite{SBurger,AndersonBP} and bright
\cite{Hulet,Salomon,Oberthaler1} one-dimensional (1D) solitons, weakly
unstable (therefore, observable) 2D solitons of the Townes type \cite%
{Townes1,Townes2}, stable quantum droplets in the 2D \cite%
{Tarruell1,Tarruell2} and 3D \cite{Inguscio}, and other dynamical modes \cite%
{Oberthaler2,CLee,WCSyu,Davidson,CKong}. The cubic self-attractive
(mean-field) nonlinearity readily predicts solitons in the 2D and 3D free
space, but these solitons are rendered completely unstable by the critical
(2D) and supercritical (3D) collapse which takes place in the same settings
\cite{LBerge,CSulem,Fibich,MalomedBA,MalomedBA2} (unless the effect of the
quantum fluctuations manifests itself, providing the collapse-suppressing
quartic self-interaction \cite{Petrov,Petrov2,Z. Luo}, which stabilizes the
quantum droplets \cite{Tarruell1,Tarruell2,Inguscio}). The instability gives
rise to the challenging problem of the prediction and creation of \emph{%
stable} 2D and 3D matter-wave solitons, including those \ with embedded
vorticity, that may be maintained by the mean-field nonlinear interactions
in appropriately modified physical settings \cite{MalomedBA2}.

Various methods have been proposed to achieve this purpose. On the one hand,
it has been predicted that long-range interactions in BEC of dipole atoms,
carrying permanent magnetic moments, can support stable 2D or even 3D
matter-wave solitons \cite{PPedri,RNath,ITikhonenkov1,MRaghunandan,JHuang}.
Another possibility is to use the combination of the spin-orbit coupling
(SOC) between two components of binary BEC and attractive mean-field
nonlinearity \cite{Sakaguchi,JiangX,LiaoB,GautamS,ChenG,ZhangY,ZhangY2,Y.
Li2}. In this connection, it is relevant to mention that SOC has been
experimentally realized in a 2D setup \cite{2D SOC}. However, the
stabilization provided by these systems is limited to the fundamental
(ground state, GS) solitons. In particular, the SOC system makes it possible
to find 2D excited-state (ES) solitons, which, however, are completely
unstable \cite{Sakaguchi}. Naturally, ES solitons exhibit a richer amplitude
and phase structure than their GS counterparts, which makes stabilization of
ES solitons another relevant problem, that has not been resolved in previous
works \cite{MalomedBA2,X. Chen}.

The objective of this work is to predict \emph{stable }2D\ solitons, \emph{%
including} ones of the ES type, in the spin-orbit-coupled binary BEC with
opposite signs (attraction and repulsion) of the self-interactions in the
two components. This arrangement, which was not addressed previously, may be
realized in the experiment by means of the Feshbach-resonance technique,
applied to the two-component BEC \cite{FR1,SBPapp,ZhangP,FR2}. The analysis
produces solitons of the semi-vortex (SV) type, defined as in Ref. \cite%
{Sakaguchi}: these are GSs with the vorticity set $\left( s_{+},s_{-}\right)
=\left( 0,1\right) $ of their two components, and ESs with extra vorticity $%
S_{+}=1,2,3,...$ added to each component, i.e., with\ the set
\begin{equation}
\left\{ s_{+},s_{-}\right\} =\left\{ S_{+},S_{+}+1\right\} .  \label{S+1}
\end{equation}%
In the experiment, SVs (alias half-vortices \cite{half}) can be induced by
vorticity-carrying laser beams resonantly coupled to particular components
of the binary BEC. Our analysis demonstrates that the ES solitons may be
stable, at least, up to $S_{+}=6$. Systematically collected numerical
results reveal that the ES solitons with the vorticity in both components
exist provided that their total does not exceed a collapse threshold $N_{%
\mathrm{thr}}$, which surpasses the well-known threshold ($N_{c}\approx 5.85$%
) for the fundamental (Townes) single-component solitons in the free 2D
space \cite{Townes,LBerge}. The value $N_{\mathrm{thr}}$ increases with the
growth of the additional vorticity $S_{+}$. These findings offer, for the
first time, a solution of the problem of the creation of stable ES solitons
in SOC systems.

The following presentation is structured as follows. In Section II, we
introduce the model, based on a system of two spin-orbit-coupled
Gross-Pitaevskii equations (GPEs), with the opposite signs of the cubic
nonlinearity in the two components. In Section III, we produce SV soliton
solutions of the GS and ES types and test their stability by means of
numerical methods. In Section IV, we address moving SV solitons and their
stability, which is a nontrivial problem, as the SOC breaks the system's
Galilean invariance. A critical velocity $V_{\mathrm{cr}}$, up to which the
stable moving solitons exist, is identified as a function of the total norm $%
N$. Finally, Section V concludes the paper.

\section{The model}

In its scaled form, the 2D system of coupled GPEs for the spinor wave
function, $\Psi =\left\{ \Psi _{+},\Psi _{-}\right\} $, with the SOC of the
Rashba type, is written as
\begin{eqnarray}
i\frac{\partial \Psi _{+}}{\partial t} &=&-\frac{1}{2}\nabla ^{2}\Psi
_{+}+\left( \frac{\partial }{\partial x}-i\frac{\partial }{\partial y}%
\right) \Psi _{-}  \notag \\
&-&\left( g_{+}|\Psi _{+}|^{2}+\gamma |\Psi _{-}|^{2}\right) \Psi _{+},
\label{+} \\
i\frac{\partial \Psi _{-}}{\partial t} &=&-\frac{1}{2}\nabla ^{2}\Psi
_{-}-\left( \frac{\partial }{\partial x}+i\frac{\partial }{\partial y}%
\right) \Psi _{+}  \notag \\
&-&\left( -g_{-}|\Psi _{-}|^{2}+\gamma |\Psi _{+}|^{2}\right) \Psi _{-},
\label{-}
\end{eqnarray}%
where positive $g_{+}$ and $g_{-}$ coefficients are strengths of the
self-attraction and repulsion in the components $\Psi _{+}$ and $\Psi _{-}$,
respectively, and $\gamma $ is the cross-interaction strength. Note that the
form in which the equations are written, with the fixed value of the SOC
coefficient, does not admit rescaling of the coordinates. Rescaling of the
wave function is possible, which may be used to fix that any of coefficients
$g_{\pm }$ and $\gamma $ is $1$. In fact, in most cases, we use this degree
of freedom to set $\gamma =1$, except for the solutions presented in Figs. %
\ref{fig2}(a,b,c), where $\gamma $ is varied, while the nomalization is
fixed by setting $g_{\pm }=1$.

The linearized form of Eqs. (\ref{+}) and (\ref{-}) for plane-wave
solutions, $\Psi _{\pm }\sim \exp \left( -i\mu _{\mathrm{linear}%
}t+ik_{x}x+ik_{y}y\right) $, yields the dispersion relation%
\begin{equation}
\mu _{\mathrm{linear}}=\pm k+k^{2}/2,  \label{mu}
\end{equation}%
where $k^{2}=k_{x}^{2}+k_{y}^{2}$. An obvious corollary of Eq. (\ref{mu}) is
that solitons may exist in the semi-infinite bandgap,
\begin{equation}
\mu <\mu _{\mathrm{cutoff}}=-1/2,  \label{cutoff}
\end{equation}%
in which expression (\ref{mu}) cannot take its value while the wavenumber
varies in the region of $0\leq k<\infty $.

Solutions to Eqs. (\ref{+}) and (\ref{-}) for solitons of the SV type, with
a real chemical potential $\mu <0$ and the set of vorticities (\ref{S+1}) in
its components, can be looked for as (cf. Ref. \cite{Sakaguchi})%
\begin{eqnarray}
\Psi _{+} &=&\exp \left( -i\mu t+iS_{+}\theta \right) r^{S_{+}}\Phi
_{+}\left( r^{2}\right) ,  \notag \\
~\Psi _{-} &=&\exp \left( -i\mu t+i\left( 1+S_{+}\right) \theta \right)
r^{1+S_{+}}\Phi _{-}\left( r^{2}\right) .  \label{ansatz}
\end{eqnarray}%
The substitution of ansatz (\ref{ansatz}) in Eqs. (\ref{+}) and (\ref{-})
leads to system of ordinary differential equations for real functions $\Phi
_{\pm }\left( r^{2}\right) $, cf. Ref. \cite{Sakaguchi}:%
\begin{gather}
\mu \Phi _{+}=-2\left[ r^{2}\Phi _{+}^{\prime \prime }+\left( 1+S_{+}\right)
\Phi _{+}^{\prime }\right]  \notag \\
+2\left[ r^{2}\Phi _{-}^{\prime }+\left( 1+S_{+}\right) \Phi _{-}\right]
\notag \\
-\left[ g_{+}\left( r^{2}\right) ^{S_{+}}\Phi _{+}^{2}+\gamma \left(
r^{2}\right) ^{1+S_{+}}\Phi _{-}^{2}\right] \Phi _{+},  \label{radial+}
\end{gather}%
\begin{gather}
\mu \Phi _{-}=-2\left[ r^{2}\Phi _{-}^{\prime \prime }+\left( 2+S_{+}\right)
\Phi _{-}^{\prime }\right] -2\Phi _{+}^{\prime }  \notag \\
-\left[ -g_{-}\left( r^{2}\right) ^{1+S_{+}}\Phi _{-}^{2}+\gamma \left(
r^{2}\right) ^{S_{+}}\Phi _{+}^{2}\right] \Phi _{-},  \label{radial-}
\end{gather}%
where $\Phi _{\pm }^{\prime }\equiv d\Phi _{\pm }/d\left( r^{2}\right) $.

To establish a relationship between the size and structure of the SV soliton
and the magnitude of the cross- and self-interaction strengths, we define
the effective radius of the soliton and the norm ratio of the two
components, respectively, as follows:
\begin{equation}
R=\left( \frac{\int r^{2}n(\mathbf{r})\mathrm{d}\mathbf{r}}{\int n(\mathbf{r}%
)\mathrm{d}\mathbf{r}}\right) ^{1/2},~F=\frac{N_{-}}{N_{+}},  \label{RF}
\end{equation}%
where $n(r)=|\Psi _{+}(r)|^{2}+|\Psi _{-}(r)|^{2}$ is the total density of
the solution, the total and component norms being%
\begin{equation}
N=\int n(\mathbf{r})d\mathbf{r\equiv }N_{+}+N_{-}\mathbf{.}  \label{N}
\end{equation}%
Obviously, $N$ is a dynamical invariant of the system of Eqs.\ (\ref{+}) and
(\ref{-}). In all cases, when the scaling is used to set either $\gamma =1$
or $g_{\pm }=1$, as said above, $N$ remains a free control parameter of
solution families. In the next section, we first set $g_{+}=g_{-}=1$, while
the case of $g_{+}\neq g_{-}$ is considered at the section.

The system also conserves the energy (Hamiltonian),%
\begin{gather}
E=\int \left[ \frac{1}{2}\left( \left\vert \nabla \Psi _{+}\right\vert
^{2}+\left\vert \nabla \Psi _{-}\right\vert ^{2}\right) \right.  \notag \\
+\Psi _{+}^{\ast }\left( \frac{\partial }{\partial x}-i\frac{\partial }{%
\partial y}\right) \Psi _{-}+\Psi _{+}\left( \frac{\partial }{\partial x}+i%
\frac{\partial }{\partial y}\right) \Psi _{-}^{\ast }  \notag \\
\left. -\frac{g_{+}}{2}\left\vert \Psi _{+}\right\vert ^{4}+\frac{g_{-}}{2}%
\left\vert \Psi _{-}\right\vert ^{4}-\gamma \left\vert \Psi _{+}\right\vert
^{2}\left\vert \Psi _{-}\right\vert ^{2}\right] \mathrm{d}\mathbf{r,}
\label{E}
\end{gather}%
with $\ast $ standing for the complex conjugate. It is relevant to mention
that, in terms of the polar coordinates $\left( r,\theta \right) $ and
spinor wave function defined as $\left\{ \tilde{\Psi}_{+}\equiv e^{i\theta
}\Psi ,\Psi _{-}\right\} $, the Hamiltonian (\ref{E}) is written as%
\begin{gather}
E=\int_{0}^{\infty }rdr\int_{0}^{2\pi }d\theta \left[ \frac{1}{2}\left(
\left\vert \nabla \Psi _{+}\right\vert ^{2}+\left\vert \nabla \Psi
_{-}\right\vert ^{2}\right) \right.  \notag \\
+\tilde{\Psi}_{+}^{\ast }\left( \frac{\partial }{\partial r}-\frac{i}{r}%
\frac{\partial }{\partial \theta }\right) \Psi _{-}+\tilde{\Psi}_{+}\left(
\frac{\partial }{\partial r}+\frac{i}{r}\frac{\partial }{\partial \theta }%
\right) \Psi _{-}^{\ast }  \notag \\
\left. -\frac{g_{+}}{2}\left\vert \Psi _{+}\right\vert ^{4}+\frac{g_{-}}{2}%
\left\vert \Psi _{-}\right\vert ^{4}-\gamma \left\vert \Psi _{+}\right\vert
^{2}\left\vert \Psi _{-}\right\vert ^{2}\right] \mathbf{.}  \label{H}
\end{gather}%
According to the Noether theorem \cite{Noether}, the invariance of
expression (\ref{H}) with respect to an arbitrary rotation, $\theta
\rightarrow \theta +\Delta \theta $, implies that the angular momentum,
\begin{gather}
M=-i\int_{0}^{\infty }rdr\int_{0}^{2\pi }d\theta \left( \tilde{\Psi}%
_{+}^{\ast }\frac{\partial }{\partial \theta }\tilde{\Psi}_{+}+\Psi
_{-}^{\ast }\frac{\partial }{\partial \theta }\Psi _{-}\right)  \notag \\
\equiv \int_{0}^{\infty }rdr\int_{0}^{2\pi }d\theta \left[ -i\left( \Psi
_{+}^{\ast }\frac{\partial }{\partial \theta }\Psi _{+}+\Psi _{-}^{\ast }%
\frac{\partial }{\partial \theta }\Psi _{-}\right) +\left\vert \Psi
_{+}\right\vert ^{2}\right] ,  \label{M}
\end{gather}%
is also a dynamical invariant. Note that the \ substitution of ansatz (\ref%
{ansatz}) in expression (\ref{M}) yields a simple relation between the
angular momentum and norm (\ref{N}) of the stationary states, which applies
equally well to the GS and ESs:%
\begin{equation}
M=\left( 1+S_{+}\right) N.  \label{MN}
\end{equation}

\section{Ground and excited states of semi-vortex solitons}

Following Refs. \cite{HuangC,ZhongR}, SV solitons can be produced in the
numerical form by means of the imaginary-time integration method \cite{Bao},
applied to Eqs. (\ref{+}) and (\ref{-}) with an input which follows the
structure of ansatz (\ref{ansatz}),%
\begin{gather}
\Psi _{+}(t=0)=A_{+}r^{S_{+}}\exp (-\alpha _{+}r^{2}+iS_{+}\theta ),  \notag
\\
\Psi _{-}(t=0)=A_{-}r^{1+S_{+}}\exp (-\alpha _{-}r^{2}+i\left(
1+S_{+}\right) \theta ),  \label{input}
\end{gather}%
where $A_{\pm }$ and $\alpha _{\pm }$ are positive real constants. As said
above, the SVs of the GS and ES types correspond, respectively, to $S_{+}=0$
and $S_{+}=1,2,3,...$, respectively. The stability of the solitons was then
identified with the help of systematic real-time simulations of their
perturbed evolution in the framework of Eqs. (\ref{+}) and (\ref{-}).

\subsection{Ground-state (GS) solitons}

A typical example of a stable solution which represents the GS of the SV
type, with norm $N=4$ and parameters $g_{+}=g_{-}=\gamma =1$ in Eqs. (\ref{+}%
) and (\ref{-}), is displayed in Fig. 1, where the densities, $\left\vert
\Psi _{\pm }\right\vert ^{2}$, demonstrate perfect axisymmetric patterns.
Parameters (\ref{RF}) for this solution are $R=1.2377$ and $F=0.5667$.

\begin{figure}[tph]
\center
\includegraphics[height=1in,width=3.2in]{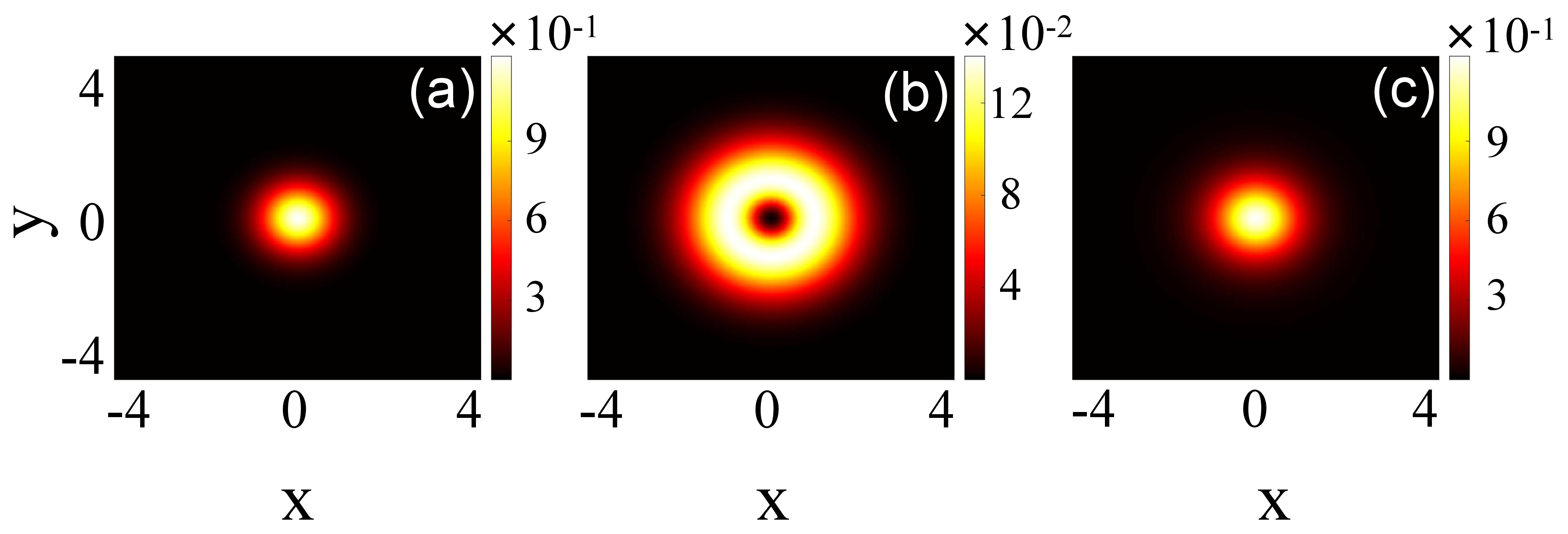}
\caption{ An example of a stable GS soliton, corresponding to ansatz (%
\protect\ref{ansatz}) with $S_{+}=0$, with the total norm $N=4$ (see Eq. (%
\protect\ref{N})), produced by the numerical solution of Eqs. (\protect\ref%
{+}) and (\protect\ref{-}) with $g_{\pm }=\protect\gamma =1$. Panels (a),
(b) and (c) display, severally, the density patterns of the zero-vorticity
and vortex components, $\left\vert \Psi _{+}\right\vert ^{2}$ and $%
\left\vert \Psi _{-}\right\vert ^{2}$, and the total density, $\left\vert
\Psi _{+}\right\vert ^{2}+\left\vert \Psi _{-}\right\vert ^{2}$.}
\label{fig1}
\end{figure}

\begin{figure}[tph]
\center
\includegraphics[height=1.8in,width=3.3in]{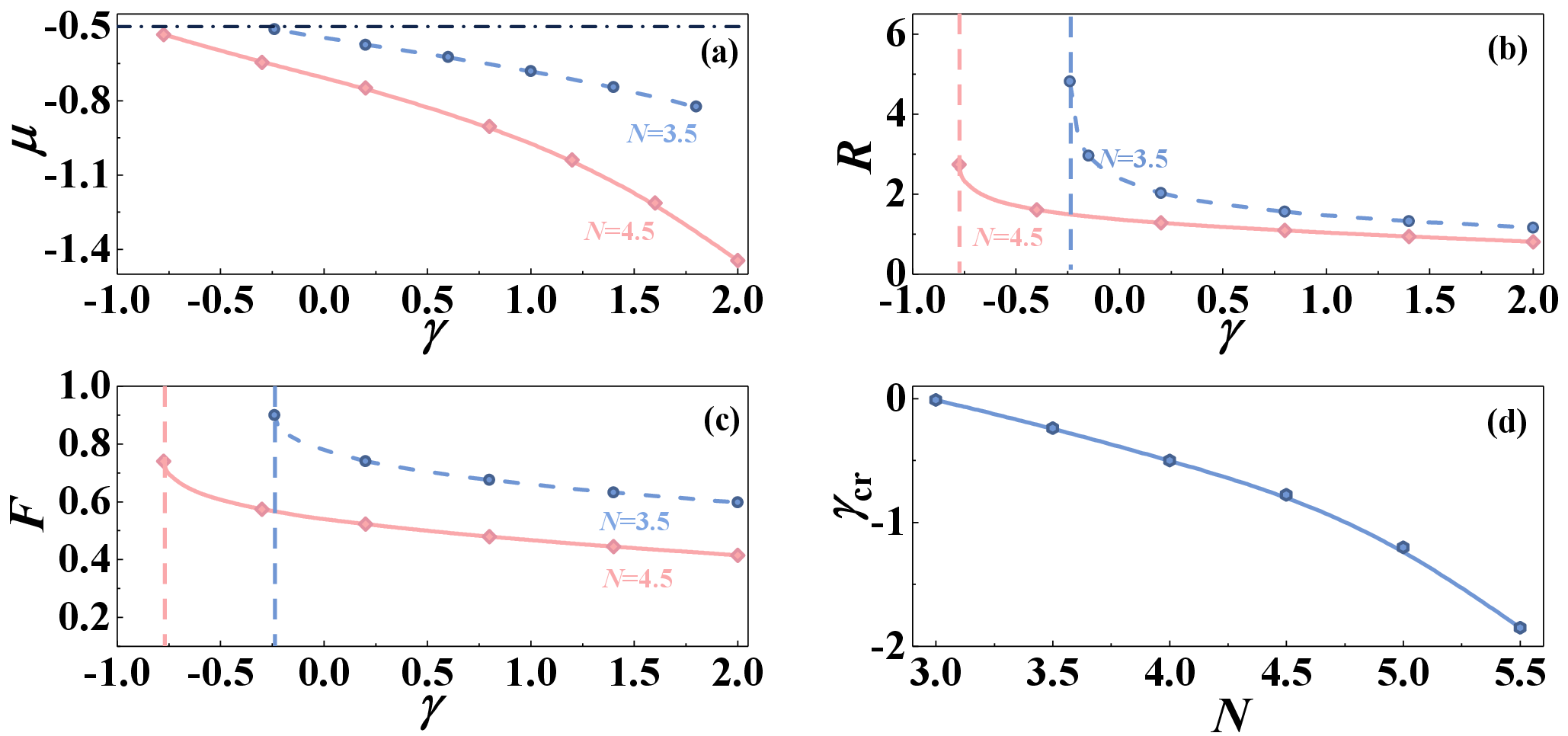}
\caption{Panels (a), (b), and (c) show, respectively, the chemical
potential, effective radius $R$, and the component-norm ratio $F$ (see Eq. (%
\protect\ref{RF})) of the ground-state SV solitons vs. the cross-interaction
strength $\protect\gamma $ for fixed values of the total norm, $N=3.5$ and $%
4.5$ (the blue dashed and red solid curves, respectively). In panel (a), the
cutoff value $\protect\mu =-1/2,$ which is the boundary of the semi-infinite
bandgap (see Eq. (\protect\ref{cutoff})), marked by the horizontal
dashed-dotten line. The respective critical values $\protect\gamma _{\mathrm{%
cr}}$ of the cross-interaction strength, at which the cutoff is attained, $%
\protect\mu \left( \protect\gamma =\protect\gamma _{\mathrm{cr}}\right)
=-1/2 $, are marked by the vertical dashed lines panels (b) and (c).
Further, $\protect\gamma _{\mathrm{cr}}$ is plotted as a function of the
norm in panel (d). In this setting, we fix $g_{\pm } =1$ in Eqs. (\protect
\ref{+}) and (\protect\ref{-}).}
\label{fig2}
\end{figure}

For two typical values of the total norm, $N=4.5$ and $N=3.5$, properties of
the GS solutions are summarized in Figs. \ref{fig2}(a,b,c), where the
chemical potential, effective radius, and the norm ratio of the two
components, defined as per Eq. (\ref{RF}), are plotted vs. the
cross-interaction strength $\gamma $ in Eqs. (\ref{+}) and (\ref{-}). In
these plots, $\gamma $ takes both positive and negative values, which
correspond, respectively, to the cross-attraction and repulsion, while the
self-interaction coefficients are fixed as $g_{+}=g_{-}=1$. In Fig. \ref%
{fig2}(a), the SV soliton suffers delocalization, turning into an infinitely
broad state with an infinitely small amplitude, as chemical potential $\mu $
is reaching the limit value (\ref{cutoff}), $\mu _{\mathrm{cutoff}}=-1/2$,
at $\gamma \rightarrow \gamma _{\mathrm{cr}}\left( N=4.5\right) =-0.775$ or $%
\gamma _{\mathrm{cr}}\left( N=3.5\right) =-0.24$. For other values of the
norm (in the interval of $3<N<5.5$), the dependence of $\gamma _{\mathrm{cr}%
} $ on $N$ is plotted in Fig. \ref{fig2}(d). Self-trapped bound states
(solitons) do not exist if $\gamma $ is \textquotedblleft too negative",
taking values $\gamma <\gamma _{\mathrm{cr}}(N)$, as the self-trapping is
obviously impossible when the overall nonlinearity is self-repulsive. The
transition to the delocalization is also demonstrated by the increase of the
effective size $R$ (defined as per Eq. (\ref{RF})) as $\gamma $ is
approaching $\gamma _{\mathrm{cr}}$.

The norm ratio $F$ of the vortical and zero-vorticity components (see Eq. (%
\ref{RF})) takes values $F<1$ because, naturally, the central hole in the
former component makes its norm smaller, cf. Figs. 1(b) and (a). The fact
that the larger part of the norm is concentrated in the self-attractive
components, $\Psi _{+}$, helps to minimize the solitons' energy and thus
stabilize them. It may also be possible to construct SV solitons with a%
\textit{\ reverse structure}, i.e., the zero-vorticity and vortical
components carried by wave functions $\Psi _{-}$ and $\Psi_{+} ,$
respectively (their ES counterparts can be introduced too). In fact, it is
easier to construct such reverse-structured SV\ solitons as solutions (\ref%
{ansatz}) with $S_{+}\geq 0$ (as adopted above), but for opposite signs of
the nonlinearity coefficients, \textit{viz}., $g_{+}<0$ and $g_{\_}<0$ in
Eqs. (\ref{+}) and (\ref{-}). While we did not analyze such reverse SV
solutions in detail, it is implausible that they may be stable, as they
place a larger share of the norm, corresponding to the zero-vorticity
component, under the action of the self-repulsion, which makes the total
energy larger, leading to destabilization. Thus, while both the GS and ES
species of the SV\ solitons keep the axial isotropy of the solutions. as is
obvious from Eqs. (\ref{ansatz}) and (\ref{input}), the solutions break the
chiral symmetry, as they favor only positive values of the angular momentum,
according to Eq. (\ref{MN}).

\subsection{ Excited-state (ES) solitons}

Stable higher-order (ES) soliton solutions have been constructed by applying
the imaginary-time-simulation method to Eqs. (\ref{+}) and (\ref{-}) with
the input taken as per Eq. (\ref{input}), up to $S_{+}=6$ (still higher
values of $S_{+}$ were not considered here, cf. Refs. \cite{HuangC,ZhongR}).
Typical examples of the so density and phase structure of the so constructed
stable ES solitons with $S_{+}=1,2,3$ and $4$ are displayed in Fig. \ref%
{fig3}, where we set $g_{\pm }=\gamma =1$ and $N=4$.

\begin{figure}[tph]
\center
\includegraphics[height=4.5in,width=5.5in]{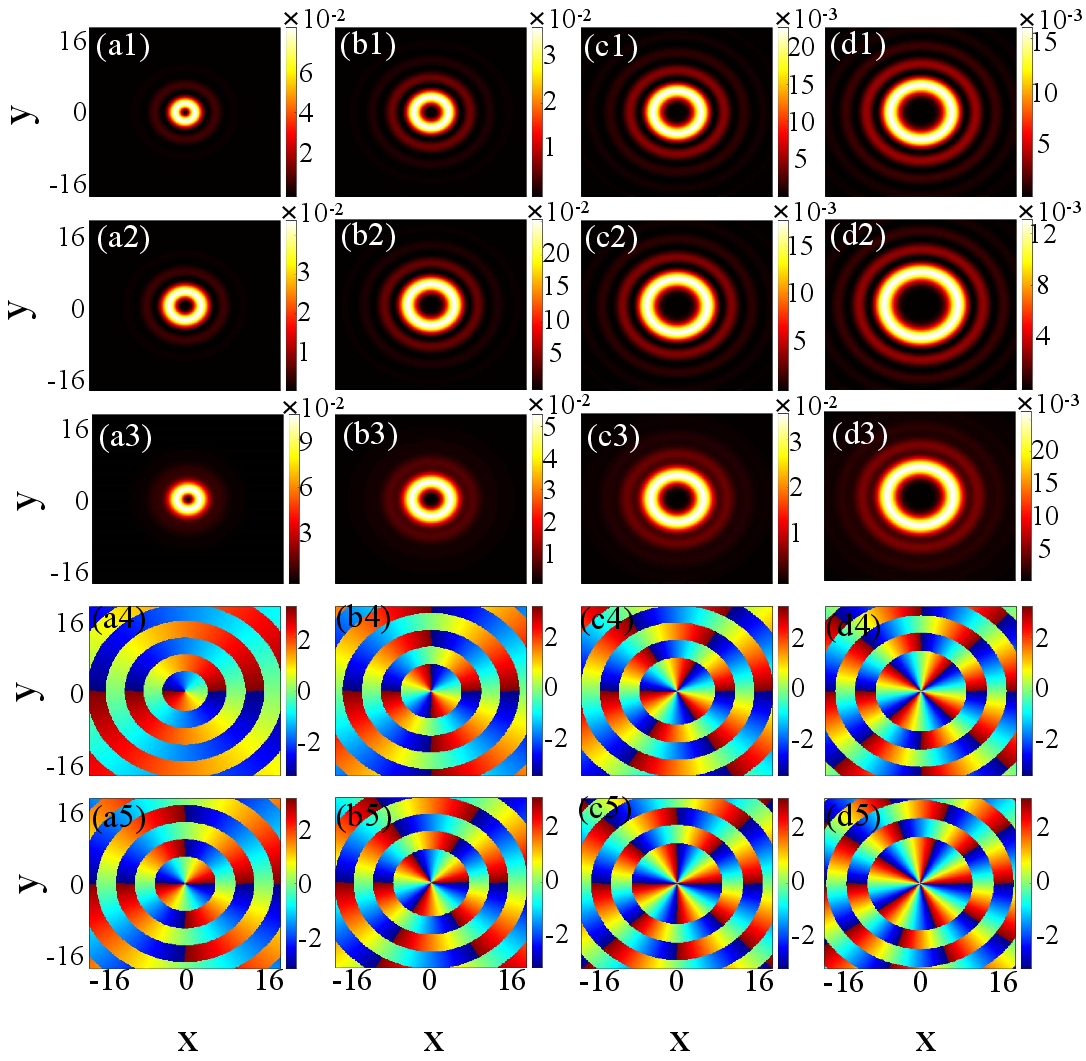}
\caption{The first and second rows: density patterns of the $\Psi _{+}$ and $%
\Psi _{-}$ components of the ES-SV solitons. The third row: the
total-density patterns, $\left\vert \Psi _{+}\right\vert ^{2}+\left\vert
\Psi _{-}\right\vert ^{2}$, for the same states. The fourth and fifth rows:
phase patterns of the components $\Psi _{+}$ and $\Psi _{-}$, respectively.
From left to right: $S_{+}=1,2,3$ and $4$. Here, the parameters are fixed as
$g_{\pm }=\protect\gamma =1$ and $N=4$. }
\label{fig3}
\end{figure}

\begin{figure}[tph]
\center
\includegraphics[height=1.8in,width=3.3in]{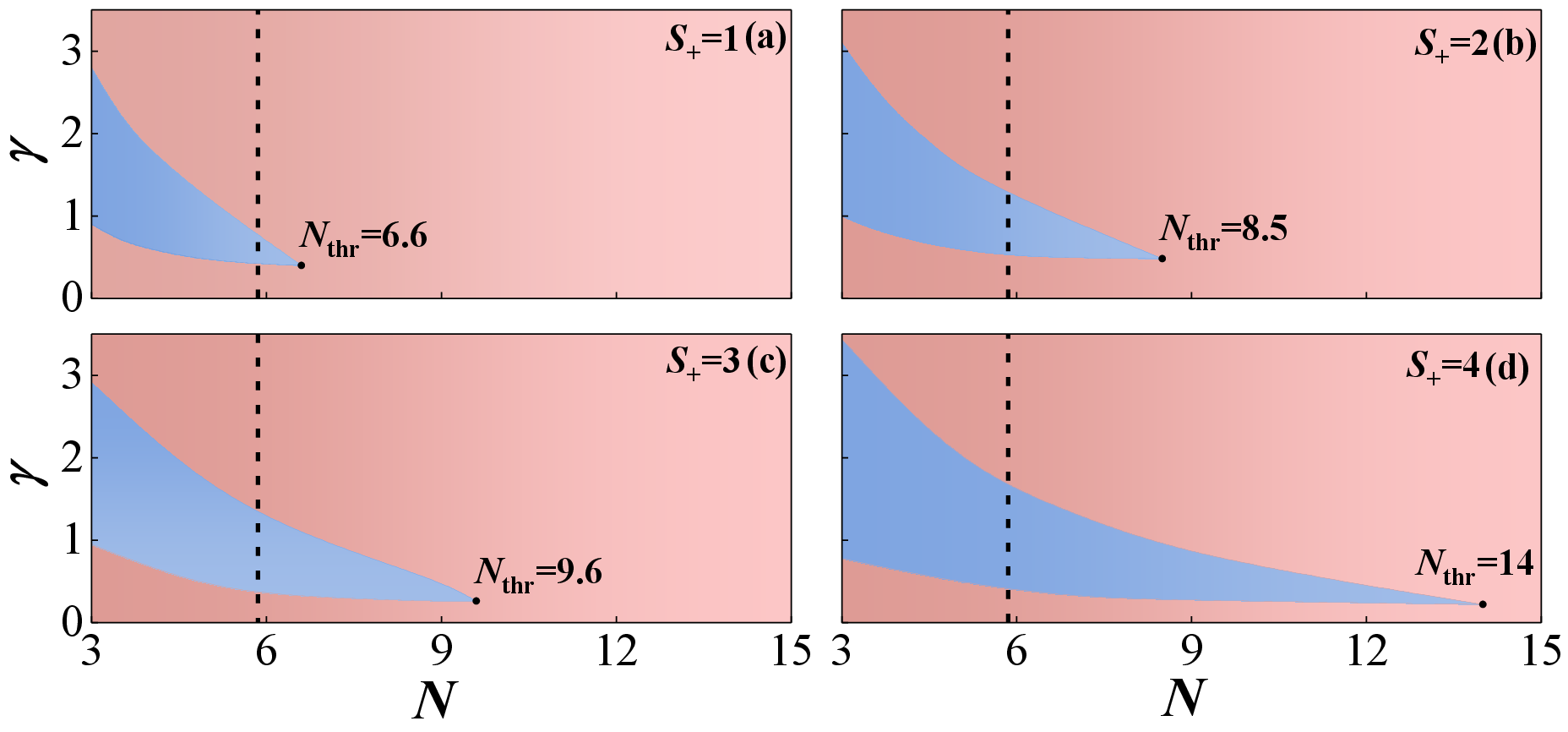}
\caption{The ES-SV solitons are stable/unstable in the blue/red areas of the
parameter plane of $(N,\protect\gamma )$ for the extra topological charges $%
S_{+}=1$ (a), $S_{+}=2$ (b), $S_{+}=3$ (c), and $S_{+}=4$ (d). Other
parameters are fixed as $g_{\pm }=1$.}
\label{fig4}
\end{figure}

The systematic numerical analysis has identified the stability charts for
the ES-SV solitons in the parameter plane of $(N,\gamma )$, which are
presented in Figs. \ref{fig4}(a)-(d) for $S_{+}=1,2,3$, and $4$ (in fact,
stable ES solitons have been found up to $S_{+}=6$, and may plausibly exist
for still larger values of $S_{+}$). The ESs are stable and unstable in the
blue and red areas, respectively. The stable ES solitons exist with the
total norm taking values below the collapse threshold $N_{\mathrm{thr}}$,
which surpasses the well-known collapse threshold ($N_{c}\approx 5.85$) for
the single-component fundamental Townes solitons in free 2D space \cite%
{LBerge}, as well as for the GS-SV solitons \cite{Sakaguchi}. As the
boundaries between stable and collapsing ES solitons, the threshold values
for the onset of the critical 2D collapse are $N_{\mathrm{thr}}=6.6$, $8.5$,
$9.6$, and $14$ for $S_{+}=1$, $2$, $3$ and $4$, respectively. Thus, the
collapse-free area expands with the increase of the extra topological
charges, $S_{+}$, or, in other words, with the increase of the soliton's
angular momentum, pursuant to Eq. (\ref{MN}). The corresponding relative
expansion of the area is $(N_{\mathrm{thr}}-N_{c})/N_{c}=12.8\%$, $45.3\%$, $%
64.1\%$ and $139\%$ for $S_{+}=1$, $2$, $3$ and $4$, respectively.

\begin{figure}[tph]
\center
\includegraphics[height=2in,width=3in]{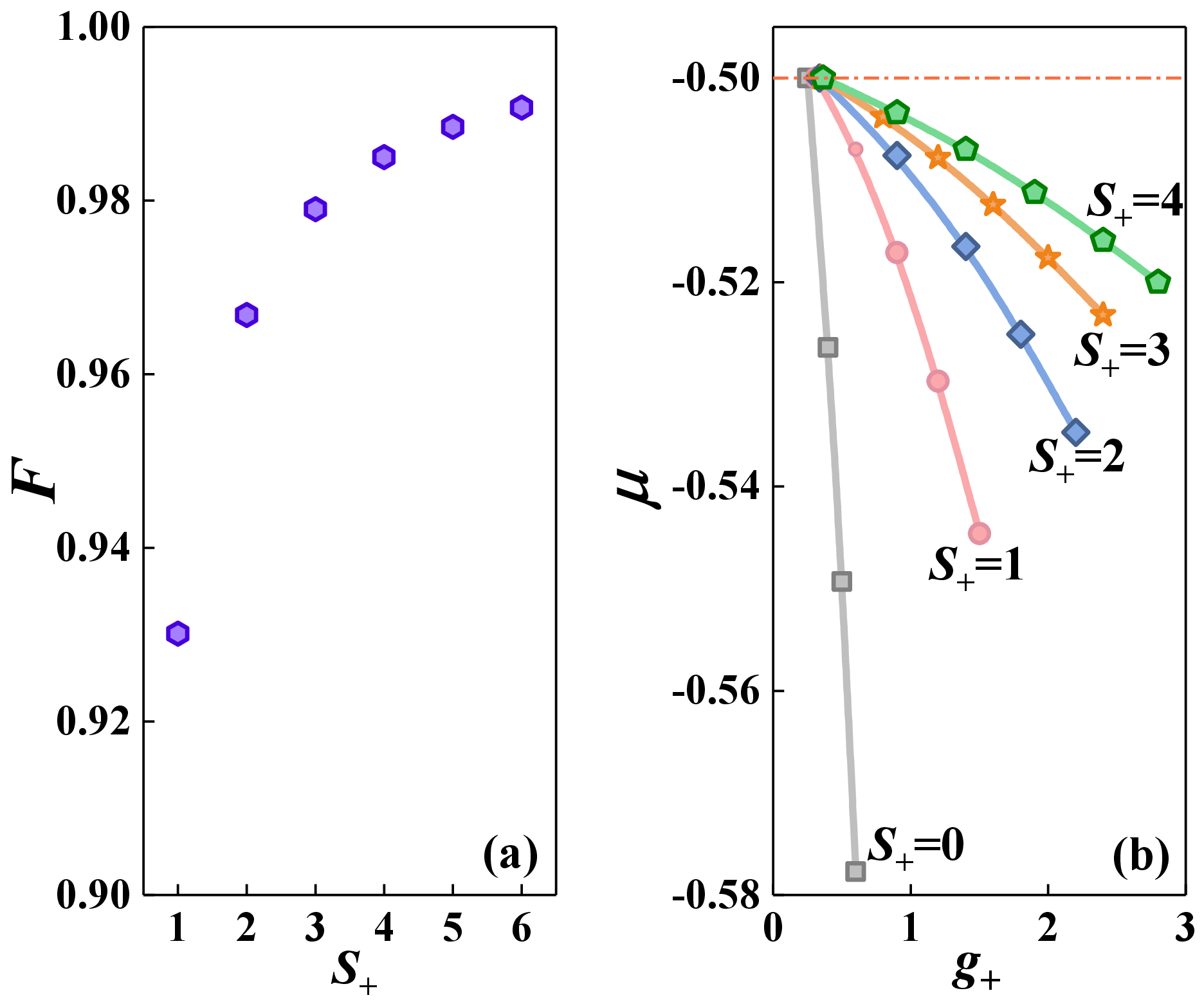}
\caption{ (a) Ratio $F=N_{-}/N_{+}$ (see Eq. (\protect\ref{RF})) for the
ES-SV solitons vs. the additional topological charge $S_{+}$, with fixed
parameters $(N,\protect\gamma,g_{+})=(4,1,1)$. (b) The solitons' chemical
potential $\protect\mu $, for different values of $S_{+}$, as a function of
the self-interaction strength $g_{+}$ in component $\Psi _{+}$, with fixed
parameters $(N,\protect\gamma,g_{-})=(4,1,1)$.}
\label{fig5}
\end{figure}

To explain the counter-intuitive phenomenon observed in Fig. \ref{fig4},
where the stability area of the ES-SV solitons expands with then increase of
the extra topological charge $S_{+}$, the corresponding ratio $F=N_{-}/N_{+}$
(see Eq. (\ref{RF})) is displayed in Fig. \ref{fig5}(a) vs. $S_{+}$ for
fixed parameters $(N,\gamma,g_{\pm })=(4,1,1)$. The growing dependence $%
F\left( S_{+}\right) $ implies the transfer of the condensate into the
self-repulsive component $\Psi _{-}$, which naturally inhibits the onset of
the collapse.

To further quantify properties of the ES-SV solitons, we fix parameters $%
(N,\gamma,g_{-})=(4,1,1)$ and vary strength $g_{+}$ of the self-attraction
component in component $\Psi _{-}$, plotting the respective dependences $\mu
\left( g_{+}\right) $ for different values of $S_{+}$ in Fig. \ref{fig5}(b).
An essential observation is that $\mu (g_{+})$ curves satisfy the
Vakhitov-Kolokolov (VK) criterion, $d\mu /dg_{+}<0$, which is a well-known
necessary stability condition for solitons maintained by a self-attractive
nonlinearity \cite{MVakhitov,Fibich} (the VK criterion takes this form for
the fixed norm, while the strength of the self-attraction varies).

\begin{figure}[tph]
\center\includegraphics[height=2in,width=2.8in]{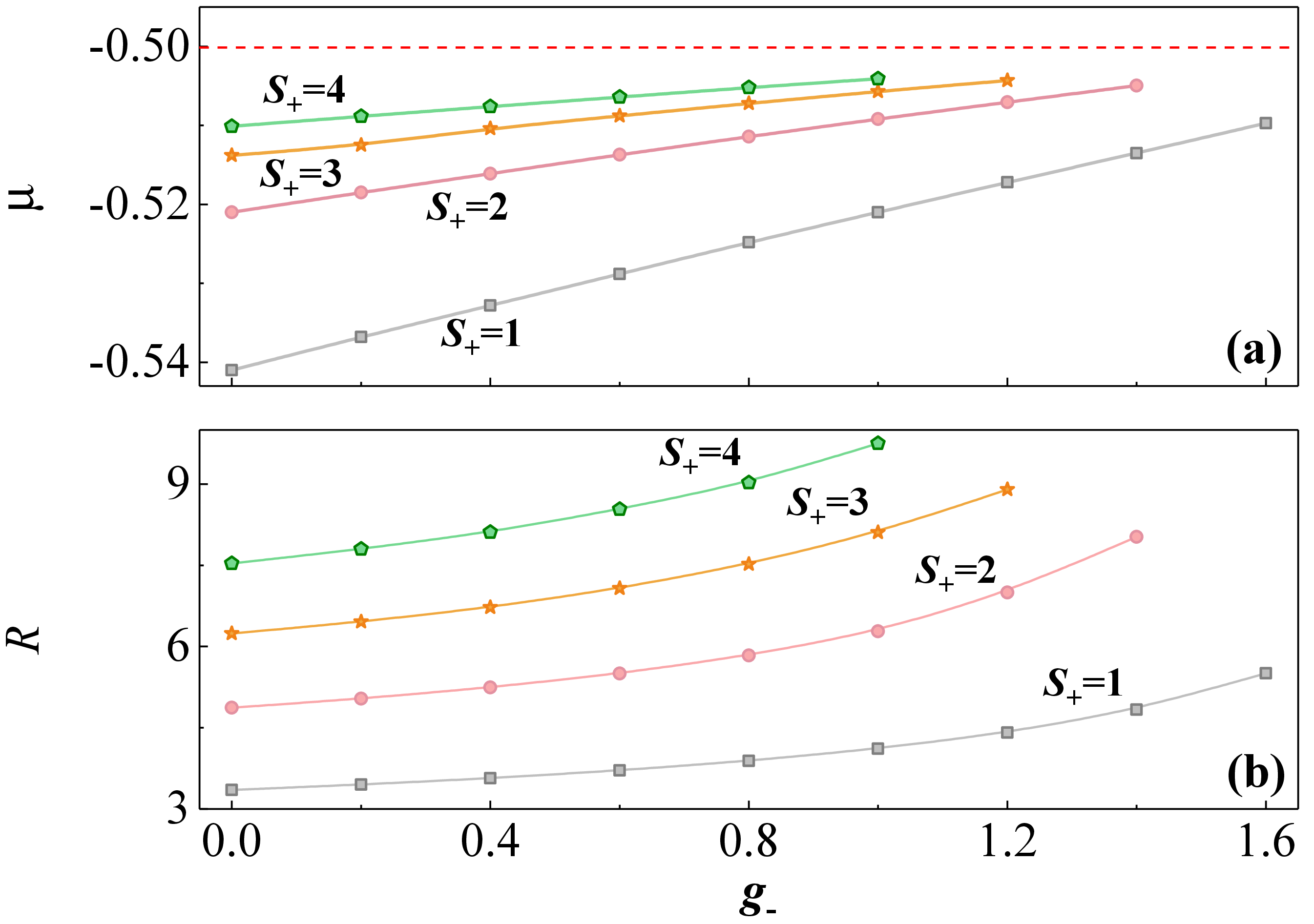}
\caption{(a,b) Chemical potential $\protect\mu $ and effective radius $R$ of
the ES-SV solitons vs. the self-repulsion strength $g_{-}$ in component $%
\Psi _{-}$ for different values of the additional topological charge $S_{+}$%
, with fixed parameters $(N,\protect\gamma,g_{+})=(4,1,1)$.}
\label{fig6}
\end{figure}

Next, Fig. \ref{fig6} displays the dependence of $\mu $ and effective
soliton's radius $R$ (see Eq. (\ref{RF})) as functions of the self-repulsion
strength $g_{-}$ in component $\Psi _{-}$ for different values of $S_{+}$,
fixing other parameters as $(N,\gamma,g_{+})=(4,1,1)$. In particular,
positive slope, $d\mu /dg_{-}>0$ of the $\mu \left( g_{-}\right) $ curves in
Fig. \ref{fig6}(a) implies that the ES-SV families satisfy the \textit{%
anti-VK criterion}, which is a necessary stability condition for solitons
with respect to the variation of the \emph{self-repulsion strength} \cite{HS}%
. In Fig. \ref{fig6}(b), the growth of the soliton's radius $R(g_{-})$ with
the increasing of $g_{-}$ demonstrates a natural trend to expansion of the
bound state under the action of stronger self-repulsion.

\section{Stability of moving 2D solitons}

The SO coupling breaks the Galilean invariance of GPEs, hence mobility of
SOC-affected solitons is a nontrivial problem. It was reported \cite%
{Sakaguchi} that, in the case of $g_{-}=-g_{+}\equiv 1$ and $\gamma <1$
(equal strengths of the self-attraction in both components of Eqs. (\ref{+})
and (\ref{-}), exceeding the strength of the inter-component attraction),
SVs can move stably, but only in a narrow interval of velocities along the $%
y $ axis, e.g., $\left\vert V_{y}\right\vert \leq V_{\mathrm{cr}}\approx
0.03 $ for the norm $N=3.7$. A much larger stability interval, e.g., $%
\left\vert V_{y}\right\vert \leq V_{\mathrm{cr}}^{(\mathrm{MM})}\approx 1.8$
for $N=3.1$ and $\gamma =2$, was found for moving \textit{mixed-mode (MM)
solitons}, which are stable in the case of $\gamma >1$, when the SV solitons
are completely unstable. The name of the MM states implies that they are
generated by inputs which feature mixtures of terms with vorticities $%
\left\{ 0,-1\right\} $ in component $\Psi _{+}$, and $\left\{ 0,+1\right\} $
in $\Psi _{-}$. Actually, solutions for GS solitons moving with velocity $%
V>V_{\mathrm{cr}}$ converge to MM states instead of the SV ones \cite%
{Sakaguchi}.

The current system produces stably moving SV solitons in much broader
velocity interval. To analyze this issue, it is natural to apply the formal
Galilean boost, with the velocity vector $\left( V_{x},V_{y}\right) $, to
Eqs. (\ref{+}) and (\ref{-}), which would keep the equations invariant in
the absence of the SOC term:
\begin{gather}
i\frac{\partial \tilde{\Psi}_{+}}{\partial t}=-\frac{1}{2}\nabla _{+}^{2}%
\tilde{\Psi}+\left( \frac{\partial }{\partial \tilde{x}}-i\frac{\partial }{%
\partial \tilde{y}}\right) \tilde{\Psi}_{-}  \notag \\
-(g_{+}|\tilde{\Psi}_{+}|^{2}+\gamma |\tilde{\Psi}_{-}|^{2})\Psi
_{+}+(iV_{x}+V_{y})\tilde{\Psi}_{-},  \label{+tilde} \\
i\frac{\partial \tilde{\Psi}_{-}}{\partial t}=-\frac{1}{2}\nabla _{-}^{2}%
\tilde{\Psi}-\left( \frac{\partial }{\partial \tilde{x}}+i\frac{\partial }{%
\partial \tilde{y}}\right) \tilde{\Psi}_{+}  \notag \\
-(-g_{-}|\tilde{\Psi}_{-}|^{2}+\gamma |\tilde{\Psi}_{+}|^{2})\tilde{\Psi}%
_{-}+(-iV_{x}+V_{y})\tilde{\Psi}_{+},  \label{-tilde}
\end{gather}%
where we define the transformed wave function and moving coordinates:
\begin{gather}
\tilde{\Psi}_{\pm }\equiv \exp \left( -\frac{i}{2}\left(
V_{x}^{2}+V_{y}^{2}\right) t+iV_{x}x+iV_{y}y\right) \Psi _{\pm }(\tilde{x},%
\tilde{y},t),  \label{tilde} \\
\tilde{x}\equiv x-V_{x}t,~y\equiv \tilde{y}-V_{y}t.  \label{xy}
\end{gather}
Solving the system of Eqs. (\ref{+tilde}) and (\ref{-tilde}) by means of the
imaginary-time-integration method, we find that the moving GS-SV solitons
remain stable in a velocity interval $V\leq V_{\mathrm{cr}}$ which is much
broader than in the system with $g_{-}=-g_{+}$, $\gamma =0$, which was
addressed in Ref. \cite{Sakaguchi}. Due to the axial isotropy of the
system's Hamiltonian, written in the form of Eq. (\ref{H}), and the
isotropic shape of the general ansatz (\ref{ansatz}) for the solutions, $V_{%
\mathrm{cr}}$ does not depend on the direction of the velocity vector.

\begin{figure}[tph]
\center
\includegraphics[height=2in,width=3in]{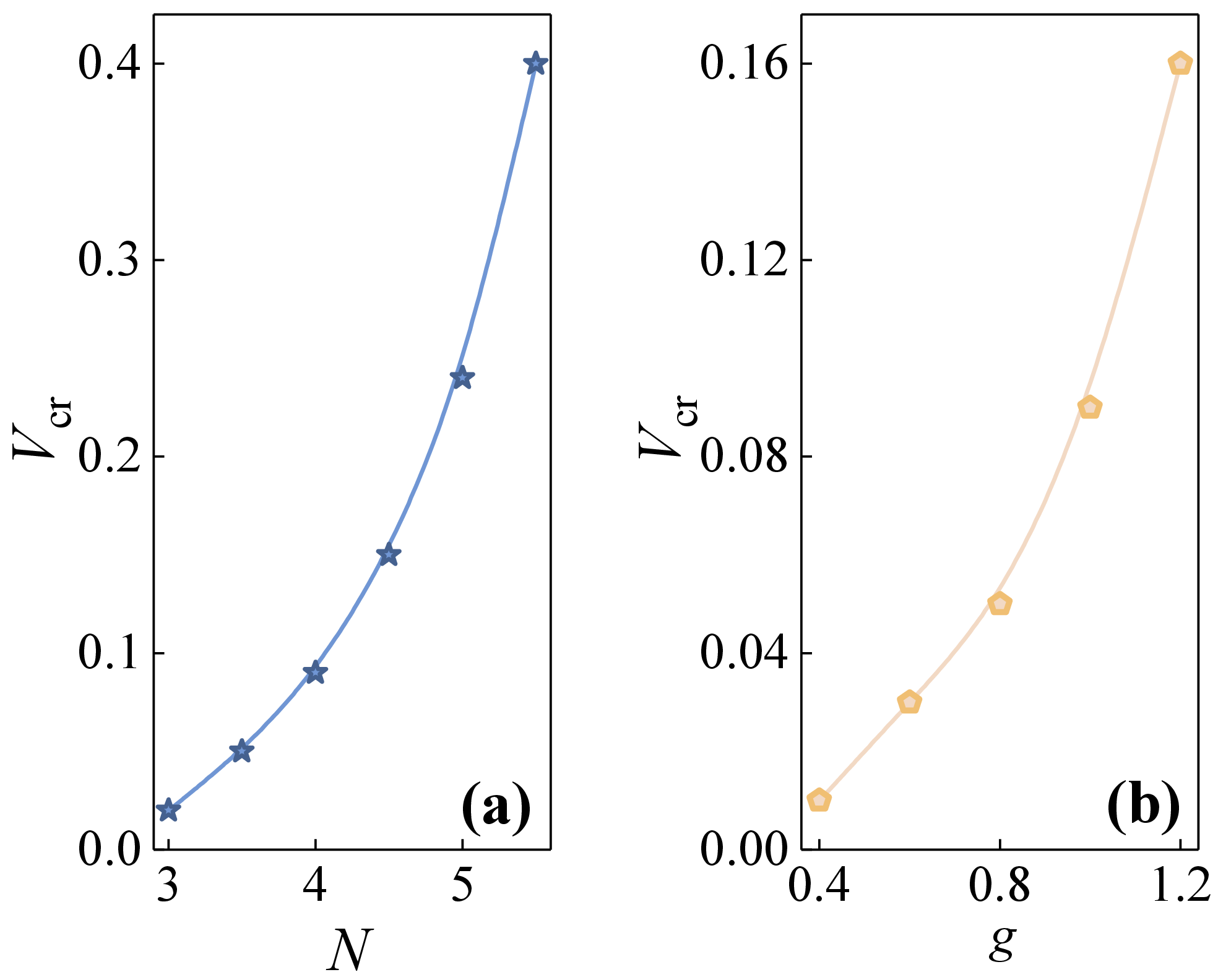}
\caption{(a) The critical velocity, $V_{\mathrm{cr}}$, up to which stably
moving GS-SV solitons are produced by the imaginary-time solution of Eqs. (%
\protect\ref{+tilde}) and (\protect\ref{-tilde}), vs. $N$ for fixed
parameters $(g_{\pm },\protect\gamma )=(1,1)$. (b) $V_{\mathrm{cr}}$ vs. $%
g_{+}=g_{-}\equiv g$ for fixed parameters $(N,\protect\gamma )=(4,1)$.}
\label{fig7}
\end{figure}

The dependences of $V_{\mathrm{cr}}$ on parameters $g_{+}=g_{-}\equiv g$ and
$N$ are summarized in Fig. \ref{fig7}. In particular, Fig. \ref{fig7}(a)
shows that, in the range of $3\leq N\leq 5.5$, the critical velocity steeply
increases with the growth of $N$. This dependence is explained by the fact
that stronger nonlinearity, which corresponds to larger norm, makes the
destabilizing effect of the velocity in Eqs. (\ref{+tilde}) and (\ref{-tilde}%
) effectively weaker. The same argument explains the growth of $V_{\mathrm{cr%
}}$ with the increase of $g$ in Fig. \ref{fig7}(b). The availability of the
stably moving SV solitons suggests considering collisions between them. In
particular, we have performed simulations for the collisions between ones
moving with opposite velocities $V_{x}=\pm0.03$ in Fig. 8. Figs. 8(a)-8(c)
display the contour plots of $\left\vert \Psi _{+}\right\vert ^{2}+
\left\vert \Psi _{-}\right\vert ^{2}$ at $t=510$, $t=1035$ and $t=1500$,
respectively. The arrows in the figures represent the direction of solitons
motion. Isosurface plot is displayed in Fig. 8(d). Clearly, the solitons
repel each other and maintain their shapes after the rebound, i.e., they
collide elastically.

As concerns the ESs with $S_{+}\geq 1$, a preliminary analysis demonstrate
that the corresponding inputs converge to the ground state, as a result of
the imaginary-time simulations of Eqs. (\ref{+tilde}) and (\ref{-tilde}). A
systematic consideration of this problem should be a subject of a separate
work. At $V>V_{\mathrm{cr}}$ the imaginary-time solution of Eqs. (\ref%
{+tilde}) and (\ref{-tilde}) demonstrates that the same input, which
produces the GS-SV solitons at $V<V_{\mathrm{cr}}$, gives rise to stable
solitons of the above-mentioned MM type, similar to what was found in Ref.
\cite{Sakaguchi}. The typical example of the mixed-mode solitons found at $%
V>V_{\mathrm{cr}}$ is displayed in Fig. 9.

\begin{figure}[tph]
\center
\includegraphics[height=3in,width=3in]{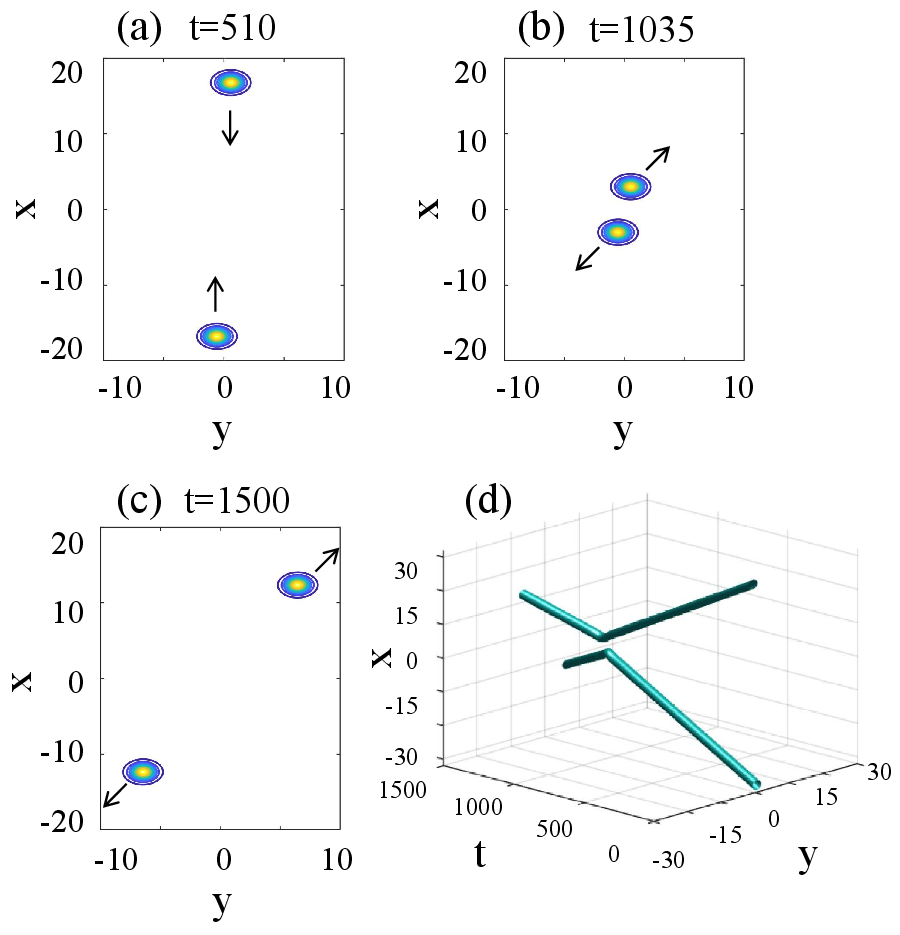}
\caption{ Collisions between SV solitons moving with opposite velocities, $%
V_{x}=\pm0.03$. Contour and isosurface plots of $\left\vert \Psi
_{+}\right\vert ^{2}+ \left\vert \Psi _{-}\right\vert ^{2}$ are displayed at
$t=510$ (a), $t=1035$ (b), $t=1500$ (c), and (d) (isosurface plot). Here,
parameters of the SV solitons are selected as $(N,\protect\gamma ,g_{\pm
},S_{+})=(4,1,1,0)$.}
\label{fig8}
\end{figure}

\begin{figure}[tph]
\center
\includegraphics[height=1.5in,width=3in]{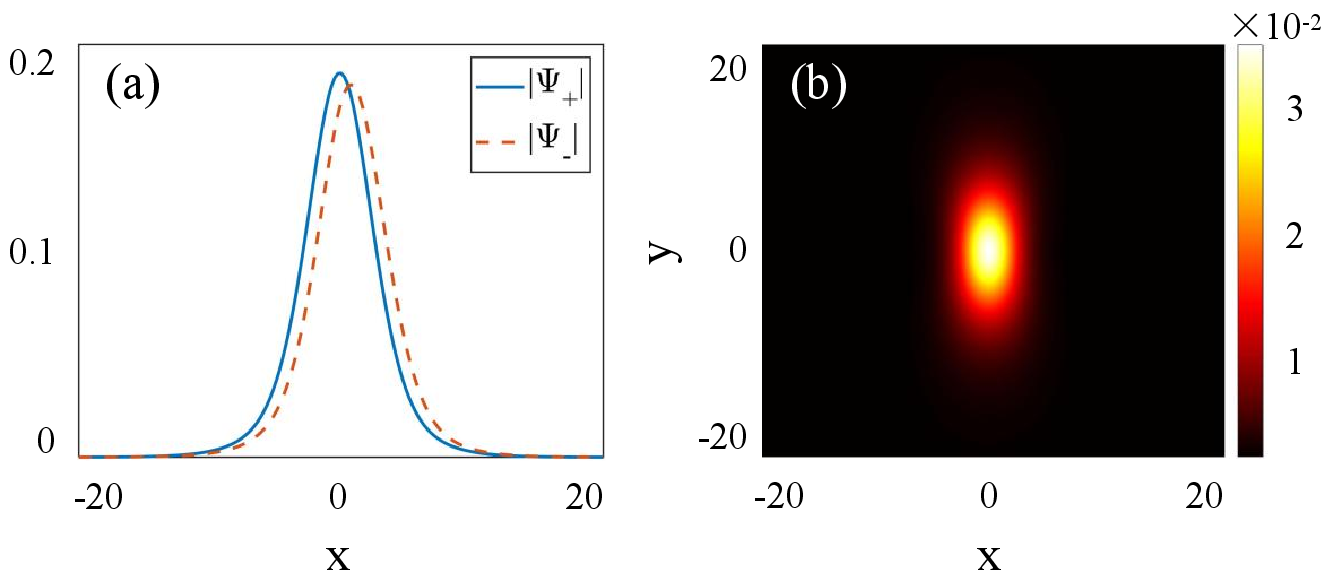}
\caption{ An example of a stable mixed mode with parameters $(N,\protect%
\gamma,g_{\pm },V_{x},V_{y})=(4,1,1,0,0.1)$. (a) One-dimensional cross
sections of the two components, $\left\vert \Psi_{+}\right\vert$ and $%
\left\vert \Psi_{-}\right\vert$. Panel (b) displays the density patterns of
vortex component, $\left\vert \Psi _{+}\right\vert ^{2}$.}
\label{fig9}
\end{figure}

\section{Conclusion}

The objective of this work is to study 2D SV (semi-vortex) solitons in the
binary BEC with two components coupled by the linear spin-orbit interaction
and nonlinear inter-component attraction. Unlike the previous work \cite%
{Sakaguchi}, we consider the system of coupled GPEs (Gross-Pitaevskii
equations) with the self-attraction and repulsion in the two components. The
most essential finding is that, in contrast to Ref. \cite{Sakaguchi}, where
the ES (excited-state) solitons, produced by the addition of the same
vorticity $S_{+}$ to both components of the GS (ground-state) soliton of the
SV type, are completely unstable in the case of the self-attraction acting
in both components, the ES-SV solitons produced by the present system have
their stability regions, expanding with the increase of $S_{+}$ (at least,
up to $S_{+}=4$). In particular, the critical norm, at which the collapse
sets in the GPE system, also increases with the growth of $S_{+}$. The study
of moving GS-SV solitons demonstrates that they exist in the interval of
velocities, $0<V<V_{\mathrm{cr}}$, which is essentially larger than the one
reported in Ref. \cite{Sakaguchi}. The critical velocity, $V_{\mathrm{cr}}$,
steeply increases with the growth of the norm and nonlinearity strength.
These findings suggest new directions for experimental studies of the
nonlinear matter-wave dynamics in the SOC (spin-orbit-coupled) BEC,
especially as concerns the challenging problem of creation of stable
solitons with high embedded vorticity.

A natural direction for further research is to extend the current analysis
for ESs of solitons of the MM (mixed-mode) type, which combine terms with
vorticities $0$ and $\pm 1$ in the two components. It may be also
interesting to explore Josephson oscillations between ESs of solitons of the
SV and MM types, cf. Ref. \cite{Z. Chen}. A challenging possibility is to
seek for (meta-) stable ESs of solitons in three-dimensional SOC BEC, cf.
Ref. \cite{ZhangY} (it is relevant to expect metastability in that case, as
the supercritical collapse cannot be suppressed in 3D, hence 3D solitons
cannot realize a ground state). Another potentially interesting direction is
the study of the stable ESs of solitons in a heteronuclear mixture \cite%
{Salasnich}, if SOC is induced in one or both species which compose the
mixture.

The underlying system of equations (\ref{+}) and (\ref{-}) does not include
a trapping potential. It is well known that 2D potentials stabilize all
trapped single-component solitons with zero vorticity, $S=0$, against the
critical collapse, and also stabilize a part of solitons with embedded
vorticity $S=1$ against the splitting. On the other hand, all the trapped
solitons with $S\geq 2$, which may be considered as counterparts of the ESs
considered here, remain unstable \cite{MalomedBA2}. In terms of the present
consideration, the inclusion of a trapping potential in Eqs. (\ref{+}) and (%
\ref{-}) may help to additionally stabilize the ESs.

It is relevant to mention that the stabilization mechanism provided by SOC
should be sufficiently stable against the action of losses, i.e., gradual
decay of the solitons' amplitude, as the SOC is, by itself, the linear
phenomenon. The topological charge of SV states provides their protection
against other external perturbations.

Recently, much interest in the experimental \cite{H. Kadau,I.
Ferrier,Tarruell1,Tarruell2,Inguscio} and theoretical \cite%
{Petrov,Petrov2,B. A. Malomed, Z. Luo,X. Zhang,Y. Li,G. Li,Y. Li1} work was
drawn to self-bound quantum droplets in BEC stabilized by quantum
fluctuations (the Lee-Huang-Yang effect \cite{LHY}). In this connection, it
will be relevant to construct stable ESs of spin-orbit-coupled quantum
droplets of the SV and MM types. To this end, it is necessary to use SOC
GPEs including the Lee-Huang-Yang corrections \cite{Y. Li}.

\section*{Acknowledgments}

The authors appreciate valuable discussions with Profs. Yongyao Li and Bin
Liu (Foshan University). This work was supported by the Project of Hunan
Provincial Education Office under Grant No. 23A0593, 23B0774, Scientific
Research Foundation of Xiangnan University for High-Level Talents, the
Applied Characteristic Disciplines of Electronic Science and Technology of
Xiangnan University (XNXY20221210), Science and Technology Innovative
Research Team in Higher Educational Institutions of Hunan Province,
Scientific research project of Xiangnan University ([2022]96), and Israel
Science Foundation (grant No. 1695/22).

\end{document}